\documentclass[aps, prstab, reprint, showpacs, amsmath, amssymb]{revtex4-1}
\usepackage[english]{babel}
\usepackage{graphicx}
\usepackage{physics}

\usepackage{color}

\begin{document}

\title{Diffraction of Cherenkov Radiation at the Open End \\ of a Shallow Corrugated Waveguide}%

\author{Sergey N.~Galyamin}
\email{s.galyamin@spbu.ru} 
\author{Aleksandra A.~Grigoreva}
\affiliation{Saint Petersburg State University, 7/9 Universitetskaya nab., St. Petersburg, 199034 Russia}

%\date{\today}%

\begin{abstract}
A problem of diffraction of a symmetrical transverse magnetic mode 
$ \text{TM}_{0l} $ 
by an open-ended cylindrical waveguide corrugated inside is considered.
A depth and a period of corrugations are supposed to be much less than the wavelength and the waveguide radius. 
Therefore a corrugated waveguide wall can be described in terms of equivalent boundary conditions, i.e. a corresponding impedance boundary condition can be applied. 
Both vacuum case and the case of uniform dielectric filling of the waveguide is considered 
The diffraction problem is solved using the modified tayloring technique in Jones formulation.
Solution of the Wiener-Hopf-Fock equation of the problem is used to obtain an infinite linear system for reflection coefficients, the latter can be solved numerically using the reduction technique. 
\end{abstract}


\maketitle
%\tableofcontents

\section{Introduction}

In recent years, the different types of corrugated structures have proved themselves as a promising terahertz devices~%
\cite{Mineo10, BanStup12, HuZhong18}. 
One of the proposed radiation mechanism is the Smith-Purcell radiation~%
\cite{Potb11, Li2006, Li2008}. 
The alternative scheme is the use of small corrugations when a generated wavelength is sufficiently larger than the corrugation parameters~%
\cite{BanStup12THz, BanStupAntipov17, Mostacci02}.
Wakefields generated by bunches moving through an infinite metallic waveguide with small corrugations were also analyzed~%
\cite{TVAA2018, GrigTVA19}. 

However, for practice it is important to investigate the diffraction process occurring when the generated wakefield incidents an open aperture of the waveguide and exits the open space.
This problem is much more complicated (compared to the problem of wakefield calculation) and is of fundamental importance.
It is worth noting that mentioned problem for the case of an open-ended waveguide with smooth perfectly conducting walls can be solved rigorously by a number of methods, with the solution having the closed form~%
\cite{Weinb, T14}.
But for applications to radiation sources this case is marginally suitable because no wakefield is generated.
For wakefield excitation, certain slow-wave structure should be added into a waveguide: wall corrugation and dielectric layer are most typical examples.  
However, such a modification complicates the procedure of rigorous solution significantly.
For example, it is know that this solution can not be obtained in the closed form~%
\cite{Mittrab}.
Therefore, problems with open-ended waveguides are typically solved using some approximate techniques~%
\cite{GTAB14, IGTT14}.
To estimate the accuracy of these methods it is extremely useful to have a rigorous solution for similar problems.

One possible way here is to solve the corresponding ``embedded'' structure first (see Refs.~%
\cite{GTVA18, GTVGA19}
 and then perform the limiting procedure~%
\cite{Mittrab}%
, but this way is rather cumbersome.
An elegant method, called the generalized tayloring technique, has been proposed several decades ago for parallel-plate waveguides with dielectric filling~%
\cite{VZh78}.
Recently, a problem of 
$ TM_{ 0 l } $
mode radiation from an open-ended circular waveguide with uniform and layered dielectric filling has been solved using the generalization of this approach to the cylindrical geometry~\cite{GVT2021,GV2021arxiv}.
In the present paper, we also utilize this approach to describe the transformation of 
$ TM_{ 0 l } $
mode at the open end of a waveguide with shallow corrugation inside and smooth outside.

Note that the mode under consideration is directly related to the Cherenkov radiation generated in the shallow corrugated waveguide. 
In conformity with the work \cite{TVAA2018}, the wave field of a bunch moving along the corrugated waveguide axis is a single-frequency wakefield. 
Besides, the connection between the electromagnetic field of a transverse magnetic eigenmode and the wave field is shown in \cite{GrigTVA19}. 
It follows from the paper \cite{GrigTVA19} that $ H_{\varphi}^W = A \cdot \text{Re} \left( \left. H_{\varphi\omega} \right|_{\omega=\omega_1} \right)$, where $H_{\varphi}^W$ is the magnetic component of the wave field, $A$ is some amplitude constant, $H_{\varphi\omega}$ is the harmonic magnetic component of the $TM_{0l}$ diffracted mode, $\omega_1$ means the wave field frequency. 
Thus, the same structure of the Cherenkov radiation field and ${TM}_{0l}$ mode field can be achieved by appropriate choosing the mode frequency and amplitude.

The term ``shallow corrugations'' means that the considered wavelength and waveguide radius are much larger than the corrugation periods.
The solution is based on using the equivalent boundary conditions (EBC)~%
\cite{Nefb, StupBane12}.
The EBC means the substitution of the corrugated surface by the smooth one with impedance boundary condition.
This approach has been successfully used for the investigation of the wavefield generated by the bunches moving through corrugated waveguides~%
\cite{TVAA2018, GrigTVA19, STG19}.
Next, to apply the aforementioned generalized tayloring technique, the reflected field in the waveguide is decomposed into a series of corrugated waveguide eigenmodes.
The field in the external area (free space) is presented, in turn, by Fourier-type integral transforms, and corresponding boundary conditions are applied to these functions (the so-called Jones formulation). 
Solution of the Wiener-Hopf-Fock equation is obtained using the factorization method~%
\cite{Weinb, Mittrab} 
and then utilized to construct an infinite linear system for reflection coefficients of waveguide modes. 

\section{Vacuum case}

\subsection{Field components}

\begin{figure}
\centering
\includegraphics[width = 0.9\linewidth]{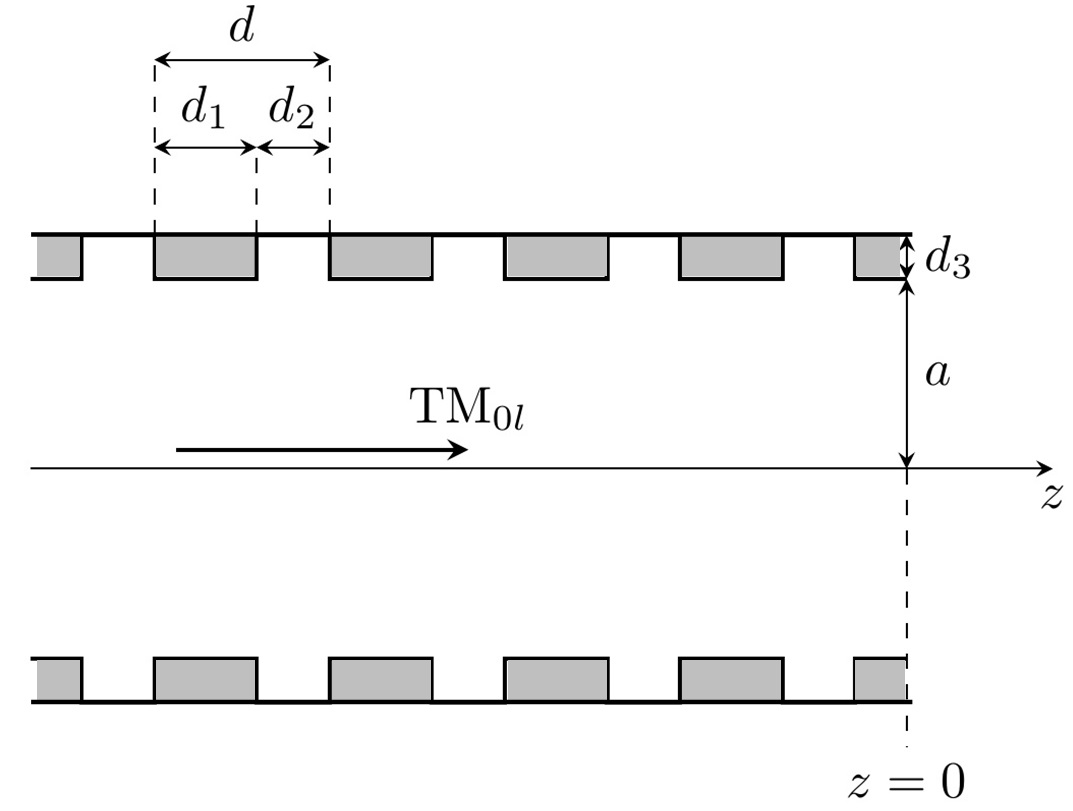}
\caption{Open-ended waveguide corrugated inside and smooth outside.}
\label{corrugated_waveguide}
\end{figure}

Geometry of the problem is shown in Fig. (\ref{corrugated_waveguide}). The harmonic $ \exp ( -i \omega t ) $ axially symmetrical transverse magnetic mode $\text{TM}_{0l}$ falling on the open end of the corrugated cylindrical waveguide. We assume that waveguide walls are perfectly conductive and the following conditions are fulfilled
\begin{equation}
d \ll a, \; d_3 \ll a, \; d \ll \lambda, d_3 \ll \lambda,
\label{cor_en}
\end{equation}
where $\lambda$ is the mode wavelength, $a$ is the waveguide radius, $d$ and $d_3$ mean corrugation period and depth respectively.

Conditions (\ref{cor_en}) allow us to replace complicated boundary conditions on the corrugated walls with equivalent boundary conditions on the smooth surface \cite{NefSivb77}
\begin{equation}
\label{cor_con}
E_{z\omega_m} \eval_{r=a} = \eta_m H_{\varphi\omega_m} \eval_{r=a},
\end{equation}
where $m$ is the mode number, $\eta_m$ means the impedance. Impedance is determined by the waveguide and mode characteristics
\begin{equation}
\label{impedance}
\eta_m ( k_{zm} ) = i\frac{\omega}{c} \left( \frac{d_2 d_3}{d} - \delta \frac{c^2 k_{zm}^2}{\omega^2} \right).
\end{equation}
Here $c$ is the speed of light in free space, $\omega$ is the mode frequency, $k_{zm}$ is the longitudinal wavenumber and $\delta$ parameters has the form
\begin{equation}
\begin{aligned}
\label{delta_imp}
\delta = d_3 + \frac{td}{2\pi} \int\limits_0^{1/\sigma} \frac{du}{ \sqrt{ ( 1 -u ) ( 1 - \sigma u ) } \left( \sqrt{1 - tu} +1 \right) } \\ + \frac{d}{2\pi} \ln \left(\frac{\sigma - 1}{\sigma} \right).
\end{aligned}\end{equation}
Parameters $t$ and $\sigma$ should be found from the following equations:
\begin{equation}\begin{aligned}
&\int\limits_0^t \frac{ \sqrt{t-u} du }{ \sqrt{ u (1-u) (\sigma-u) } } = \pi \frac{d_1}{d}, \\
&\int\limits_t^1 \frac{ \sqrt{u-t} du }{ \sqrt{u (1-u) (\sigma-u) } } = 2\pi \frac{d_3}{d}.
\end{aligned}\end{equation}

The mode structure of corrugated waveguide is considered in \cite{GrigTVA19} in detail. The incident field components can be written in form
\begin{equation}
\label{initial}
\begin{aligned}
&H_{\varphi \omega}^{\left(i\right)} = J_1 \left( \chi_{l} r \right) \exp\left( i k_{zl} z \right), \\
&E_{r\omega}^{\left(i\right)} = \frac{c}{\omega} k_{zl} J_1 \left( \chi_l r \right) \exp\left( ik_{zl} z \right), \\
&E_{z\omega}^{\left(i\right)} = \frac{ic}{\omega} \chi_l J_0 \left( \chi_l r \right) \exp\left( i k_{zl} z \right),
\end{aligned}
\end{equation}
where $J_{0,1} (\chi_l r)$ are Bessel functions, the longitudinal wavenumber is equal to 
\begin{equation}
\label{kz}
k_{ z l } = \sqrt{ \omega^2/ c^2 - \chi_l^2 },
\quad
\mathrm{ Im }  (k_{ z l }) > 0,
\end{equation}
the transverse wavenumbers $ \chi_l $ are determined by the dispersion equation
\begin{equation}
\label{disp_eq}
\frac{\omega}{c} \eta_l J_1 \left( \chi_l a \right) - i \chi_l J_0 \left( \chi_l a \right) = 0.
\end{equation}
Further we will suppose that $ \chi_l $ are known (they can be found numerically for arbitrary specific set of problem parameters).

We use the following approach to solving the problem. The reflected field is represented as a series of corrugated waveguide eigenmodes  
\begin{equation}
\label{reflected}
\begin{aligned}
&H_{\varphi \omega}^{\left(r\right)} = \sum\limits_{m=1}^{\infty} R_m J_1 \left( \chi_{m} r \right) \exp\left( -i k_{zm} z \right), \\
&E_{r\omega}^{\left(r\right)} = -\frac{c}{\omega} \sum\limits_{m=1}^{\infty} R_m k_{zm} J_1 \left( \chi_m r \right) \exp\left( -ik_{zm} z \right), \\
&E_{z\omega}^{\left(r\right)} = \frac{ic}{\omega} \sum\limits_{m=1}^{\infty} R_m \chi_m J_0 \left( \chi_m r \right) \exp\left( -i k_{zm} z \right),
\end{aligned}
\end{equation}
Mode excitation coefficients $R_m$ are unknown.
Note that expressions (\ref{initial}) and (\ref{reflected}) are written as a solution of equation for magnetic field component $H_{\varphi \omega}$
\begin{equation}
\label{eq_h}
\left( \frac{ \partial^2 }{ \partial r^2} + \frac{1}{r} \frac{ \partial  }{ \partial r } - \frac{1}{r^2} + \frac{ \partial^2 }{\partial z^2 } + \frac{\omega^2}{c^2} \right) H_{\varphi \omega} = 0. 
\end{equation}

\begin{figure}
\centering
\includegraphics[]{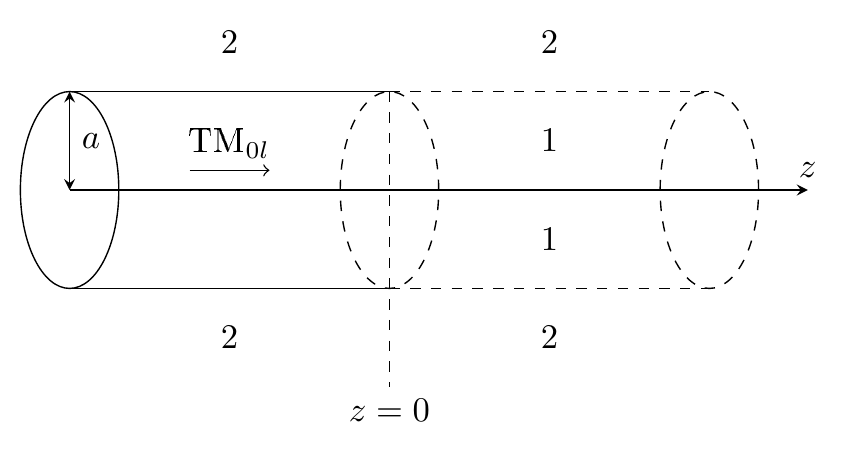}
\caption{Geometry of the problem for vacuum case and main notations. Boundary conditions~\eqref{cor_con} are applied inside the waveguide while the outside waveguide wall is smooth and perfectly conductive. Dashed lines denote the imaginary waveguide continuation dividing the outer space into subregions ``1'' and ``2''. }
\label{waveguide}
\end{figure}

In order to specify the field in the area external to the waveguide we devide area $r>a$ into two parts denoted as $``1''$ and $``2''$ (see Fig. (\ref{waveguide})). The unknown fields in areas $``1''$ and $``2''$ can be written using the following integral transform:
\begin{align}
&\Psi_-^{\left(2\right)} \left(r, \alpha\right) = \frac{1}{2\pi} \int\limits_{-\infty}^{0} H_{\varphi \omega}^{\left(2\right)} \left(r,z\right) \exp\left(i \alpha z\right) dz, \label{psi-} \\
&\Psi_+^{\left(1,2\right)} \left(r, \alpha\right) = \frac{1}{2\pi} \int\limits_{0}^{+\infty} H_{\varphi \omega}^{\left(1,2\right)} \left(r,z\right) \exp\left(i \alpha z\right) dz. \label{psi+}
\end{align}
The result of the transform (\ref{psi-}) or (\ref{psi+}) is the function regular in areas $\text{Im } \alpha < 0$ and $\text{Im } \alpha > 0 $ respectively \cite{Mittrab}. Note that function $\Psi_-^{(1)} (r, \alpha)$ is not defined since the area $``1"$ is determined only for $z>0$.

\subsection{Equation for functions $\Psi_\pm^{(2)} (r,\alpha)$}

Let consider the area $``2''$ and apply the integral transform (\ref{psi-})-(\ref{psi+}) to the equation for the magnetic field component (\ref{eq_h})
\begin{equation}\begin{aligned}
\frac{1}{2\pi} \int\limits_{-\infty}^{+\infty} \left( \frac{ \partial^2 }{ \partial r^2} + \frac{1}{r} \frac{ \partial  }{ \partial r } - \frac{1}{r^2}  + \frac{\omega^2}{c^2} \right) H_{\varphi \omega}^{\left(2\right)} e^{i \alpha z} dz \\ + \frac{1}{2\pi} \int\limits_{-\infty}^{+\infty} \frac{ \partial^2  H_{\varphi \omega}^{\left(2\right)} }{ \partial z^2} e^{i \alpha z} dz = 0.
\end{aligned}\end{equation}
By virtue of (\ref{psi-})-(\ref{psi+}) the first integral has the simple meaning
\begin{equation}\begin{aligned}
\frac{1}{2\pi} \int\limits_{-\infty}^{+\infty} \left( \frac{ \partial^2 }{ \partial r^2} + \frac{1}{r} \frac{ \partial  }{ \partial r } - \frac{1}{r^2}  + \frac{\omega^2}{c^2} \right) H_{\varphi \omega}^{\left(2\right)} e^{i \alpha z} dz \\ = \left( \frac{ \partial^2 }{ \partial r^2} + \frac{1}{r} \frac{ \partial  }{ \partial r } - \frac{1}{r^2}  + \frac{\omega^2}{c^2} \right) \left( \Psi_+^{\left(2\right)} + \Psi_-^{\left(2\right)} \right).
\label{eq_h2}
\end{aligned}\end{equation}
Integrating the second term in (\ref{eq_h2}) by parts twice we obtain the following expression
\begin{equation}
\begin{aligned}
&\int\limits_{-\infty}^{+\infty} \frac{ \partial^2  H_{\varphi \omega}^{\left(2\right)} }{ \partial z^2} e^{i \alpha z} dz  \\ 
&{=} \left. \left[  \frac{ \partial H_{\varphi \omega}^{\left(2\right)} }{ \partial z } e^{i \alpha z} {-} i \alpha H_{\varphi \omega}^{\left(2\right)} e^{i \alpha z} {-} \alpha^2 \int\limits_{-\infty}^0 H_{\varphi\omega}^{\left(2\right)} e^{i \alpha z} dz \right] \right|_{-\infty}^0 \\ 
&{+} \left. \left[  \frac{ \partial H_{\varphi \omega}^{\left(2\right)} }{ \partial z } e^{i \alpha z} {-} i \alpha H_{\varphi \omega}^{\left(2\right)} e^{i \alpha z} {-} \alpha^2 \int\limits_{0}^{+\infty} H_{\varphi\omega}^{\left(2\right)} e^{i \alpha z} dz \right] \right|_{0}^{+\infty}.
\end{aligned}
\end{equation}
It is assumed that $\left| H_{\varphi\omega}^{\left(2\right)} \right| \to 0 $ and $\abs{\partial{H_{\varphi\omega}^{\left(2\right)}}/\partial{z}} \to 0$ at $\left| z \right| \to \infty$. Furthermore, area $``2"$ doesn't contain any boundary at $z=0$ so functions $H_{\varphi\omega}^{\left(2\right)}$ and $\partial H_{\varphi\omega}^{\left(2\right)} / \partial z$ are continuos at $z=0$. Thus, the second integral in (\ref{eq_h2}) is equal to
\begin{equation}
\frac{1}{2\pi} \int\limits_{-\infty}^{+\infty} \frac{ \partial^2  H_{\varphi \omega}^{\left(2\right)} }{ \partial z^2} e^{i \alpha z} dz = -\alpha^2 \left( \Psi_-^{\left(2 \right)} + \Psi_+^{\left(2 \right)} \right).
\label{int_z_2}
\end{equation}
Combining expressions (\ref{int_z_2}) and (\ref{eq_h2}) we finally obtain the differential equation for function $\Psi^{(2)} (r, \alpha)$ 
\begin{equation}
\begin{aligned}
&\left( \frac{ \partial^2 }{ \partial r^2} + \frac{1}{r} \frac{ \partial  }{ \partial r } - \frac{1}{r^2}  + \kappa^2 \right) \Psi^{(2)} (r, \alpha) = 0, \\
&\Psi^{(2)} (r, \alpha) = \Psi_-^{(2)} (r, \alpha) + \Psi_+^{(2)} (r, \alpha),
\label{eq_psi_2}
\end{aligned}
\end{equation}
where $\kappa^2 (\alpha) = {\omega^2}/{c^2} - \alpha^2$.

The equation (\ref{eq_psi_2}) can be easily reduced to the first order Bessel equation and its solution in area $r>a$ is well-known
\begin{equation}
\Psi^{ ( 2 ) } ( r, \alpha ) = C_2 H_1^{(1)} (\kappa r).
\label{psi_2_sol}
\end{equation}
Coefficient $C_2$ will be defined later using the corresponding continuity conditions.

\subsection{Equation for function $\Psi_+^{(1)} (r,\alpha)$}

Now let consider the area $``1"$ and apply the integral transform (\ref{psi+}) to the equation for the magnetic field component (\ref{eq_h})
\begin{equation}\begin{aligned}
\frac{1}{2\pi} \int\limits_{0}^{+\infty} \left( \frac{ \partial^2 }{ \partial r^2} + \frac{1}{r} \frac{ \partial  }{ \partial r } - \frac{1}{r^2}  + \frac{\omega^2}{c^2} \right) H_{\varphi \omega}^{\left(1\right)} e^{i \alpha z} dz \\ + \frac{1}{2\pi} \int\limits_{0}^{+\infty} \frac{ \partial^2  H_{\varphi \omega}^{\left(1\right)} }{ \partial z^2} e^{i \alpha z} dz = 0.
\end{aligned}\end{equation}
Performing calculations similar to those for function $\Psi_+^{(2)} (r,\alpha)$ we obtain
\begin{equation}\begin{aligned}
\frac{1}{2\pi} \int\limits_{0}^{+\infty} \left( \frac{ \partial^2 }{ \partial r^2} + \frac{1}{r} \frac{ \partial  }{ \partial r } - \frac{1}{r^2}  + \frac{\omega^2}{c^2} \right) H_{\varphi \omega}^{\left(1\right)} e^{i \alpha z} dz \\ = \left( \frac{ \partial^2 }{ \partial r^2} + \frac{1}{r} \frac{ \partial  }{ \partial r } - \frac{1}{r^2}  + \frac{\omega^2}{c^2} \right) \Psi_+^{(1)},
\end{aligned}\end{equation}
\begin{equation}\begin{aligned}
\frac{1}{2\pi} \int\limits_{0}^{+\infty} \frac{ \partial^2  H_{\varphi \omega}^{(1)} }{ \partial z^2} e^{i \alpha z} dz = -\alpha^2 \Psi_+^{(1)} \\ + \frac{1}{2\pi} \left( i \alpha H_{\varphi\omega}^{(1)} \eval_{z=+0} - \pdv{H_{\varphi\omega}^{(1)}}{z} \eval_{z=+0} \right).
\end{aligned}\end{equation}
As a result, the following inhomogeneous differential equation can be written for function $\Psi_+^{(1)}$
\begin{align}
&\left( \frac{ \partial^2 }{ \partial r^2} + \frac{1}{r} \frac{ \partial  }{ \partial r } - \frac{1}{r^2}  + \kappa^2 \right) \Psi_+^{(1)} = F^{(1)}, \label{eq_psi_1} \\
&F^{(1)} = \frac{1}{2\pi} \left( \pdv{H_{\varphi\omega}^{(1)}}{z} \eval_{z=+0} - i \alpha H_{\varphi\omega}^{(1)} \eval_{z=+0} \right). \label{F1}
\end{align}

Explicit form of function $F^{(1)}$ can be found using the continuity condition for field components $H_{\varphi\omega}$ and $E_{r\omega}$ at surface $z=0$, $r<a$. 
The continuity of function $H_{\varphi\omega}$ means
\begin{equation}
\left(H_{\varphi\omega}^{(i)} + H_{\varphi\omega}^{(r)} \right) \eval_{z=-0} = H_{\varphi\omega}^{(1)} \eval_{z=+0}.
\end{equation}
The substitution of the incident and reflected field by expression (\ref{initial}) and (\ref{reflected}) leads to the following equality
\begin{equation}
H_{\varphi\omega}^{(1)} \eval_{z=+0} = J_1 ( \chi_l r ) + \sum\limits_{m=1}^{\infty} R_m J_1 ( \chi_m r ).
\label{h_cont}
\end{equation}
In order to use the continuity of function $E_{r\omega}$ we rewrite this component in accordance with Maxwell equations
\begin{equation}
E_{r\omega} = -\frac{ic}{\omega} \pdv{ ( H_{\varphi\omega} ) }{z}.
\label{ez_hf}
\end{equation}
Thus, the continuity of function $E_{r\omega}$ leads to the following:
\begin{equation}
\left( \pdv{ H_{\varphi\omega}^{(i)} }{z} + \pdv{ H_{\varphi\omega}^{(r)} }{z} \right) \eval_{z=-0} = \pdv{ H_{\varphi\omega}^{(1)} }{z} \eval_{z=+0}. 
\end{equation}
Using (\ref{initial})-(\ref{reflected}), one can obtain
\begin{equation}
\pdv{ H_{\varphi\omega}^{(1)} }{z} \eval_{z=+0} = i k_{zl} J_1 ( \chi_l r ) - i \sum\limits_{m=1}^{\infty} R_m k_{zm} J_1 ( \chi_m r ).
\label{e_cont}
\end{equation}

The final form of function $F^{(1)}$ is obtained by substitution of (\ref{h_cont}) and (\ref{e_cont}) in (\ref{F1})
\begin{equation}\begin{aligned}
F^{(1)} (r,\alpha) = \frac{i}{2\pi} (k_{zl} - \alpha)J_1 ( \chi_l r ) \\ - \frac{i}{2\pi} \sum\limits_{m=1}^{\infty} R_m ( k_{zm} + \alpha ) J_1 ( \chi_m r ).
\label{F1}
\end{aligned}\end{equation}

The solution of equation (\ref{eq_psi_1}) is found as a sum of general solution of a homogeneous equation $\Psi_{\text{g.s.}}^{(1)} (r,\alpha)$ and particular solution of an inhomogeneous equation $\Psi_p^{(1)} (r,\alpha)$. Since the homogeneous equation (\ref{eq_psi_1}) can be reduced to the first-order Bessel equation general solution has the simple form
\begin{equation}
\Psi_{\text{g.s.}}^{(1)} (r,\alpha) = C_1 J_1 ( \kappa r ).
\end{equation}
Based on the form of the equation (\ref{eq_psi_1}) right part, we present the particular solution as a series of Bessel functions 
\begin{equation}
\label{part_sol}
\Psi_p^{(1)} (r, \alpha) = A J_1 ( \chi_l r ) + \sum\limits_{m=1}^{\infty} B_m J_1 ( \chi_m r ).
\end{equation}
Substituting this form in (\ref{eq_psi_1}), one can obtain
\begin{equation}\begin{aligned}
\label{part_sol2}
( \kappa^2 - \chi_l^2 ) A J_1 ( \chi_l r ) + \sum\limits_{m=1}^{\infty} B_m ( \kappa^2 - \chi_m^2 ) J_1 ( \chi_m r ) \\ = \frac{i}{2\pi} ( k_{zl} - \alpha ) J_1 ( \chi_l r ) - \frac{i}{2\pi} \sum\limits_{m=1}^{\infty} R_m ( k_{zm} + \alpha ) J_1 ( \chi_m r ).
\end{aligned}\end{equation}
To determine coefficients $A$, $\qty{B_m}$ we multiply the equation (\ref{part_sol2}) by function $r \overline{J_1 ( \chi_p r )}$ (the bar means the complex conjugation) and integrate over the radial variable $r$. Then the orthogonal property of Bessel functions \cite{GrigTVA19} is used
\begin{equation}\begin{aligned}
\int\limits_0^a \overline{J_1 ( \chi_p r )} J_1 ( \chi_n r ) r dr = \delta_{np} \left[ \frac{a^2}{2} ( J_1^2 ( \chi_n a ) + J_0^2 ( \chi_n a ) ) \right. \\ \left. - \frac{a}{\chi_n} J_0 ( \chi_n a ) J_1 ( \chi_n a ) \right],
\label{orth_prop}
\end{aligned}\end{equation}
where $\delta_{np}$ means the Kronecker symbol. These mathematical transformations lead to the following result
\begin{equation}
\begin{aligned}
&A = \frac{ i }{ 2 \pi } \frac{ k_{ z l } - \alpha }{ k_{ z l }^2 - \alpha^2 } = \frac{ i }{ 2 \pi } \frac{ 1 }{ k_{ z l } + \alpha }, \\ 
&B_m = -\frac{ i }{ 2 \pi } R_m \frac{ k_{ z m } + \alpha }{ k_{ z m }^2 - \alpha^2 } =
-\frac{ i }{ 2 \pi } \frac{ R_m }{ k_{ z m } - \alpha },
\end{aligned}
\end{equation}
where $ k_{ z m } $ is defined by \eqref{kz}.

So, the solution of equation (\ref{eq_psi_1}) has the form 
\begin{equation}\begin{aligned}
\label{psi_1_sol}
\Psi_+^{(1)} (r,\alpha) = C_1 J_1 ( \kappa r ) + \frac{i}{2\pi} \frac{ J_1 ( \chi_l r ) }{  k_{ z l } + \alpha } \\  - \frac{i}{2\pi} \sum\limits_{m=1}^{\infty} R_m \frac{ J_1 ( \chi_m r ) }{ k_{zm} - \alpha } .
\end{aligned}\end{equation}

\subsection{Functions $\Phi_{\pm}^{(1,2)} (r,\alpha)$}

Introduce new function
\begin{equation}
\begin{aligned}
\label{Phi1_def}
\Phi_+^{ ( 1 ) } ( r, \alpha ) 
= 
\frac{ 1 }{ 2 \pi } 
\int\limits_{ 0 }^{ +\infty }
E_{ z \omega }^{ ( 1 ) } ( r, z )  
\exp( i \alpha z)  dz,
\end{aligned}
\end{equation}
\begin{equation}\begin{aligned}
\label{Phi2_def}
\Phi^{ ( 2 ) } ( r, \alpha ) 
= 
\frac{ 1 }{ 2 \pi } 
\int\limits_{- \infty }^{ \infty }
E_{ z \omega }^{ ( 2 ) } ( r, z ) 
\exp \left( i \alpha z \right) dz = \\
= \Phi_+^{ ( 2 ) } (r,\alpha) + \Phi_-^{ ( 2 ) } (r, \alpha).
\end{aligned}\end{equation}
%
%where coefficient $\eta (\alpha)$ is defined by analogy of the impedance \eqref{impedance}
%%
%\begin{equation}
%\label{eta_alpha}
%\eta (\alpha) = \frac{i \omega}{c} \left( \frac{d_2 d_3}{d} - \delta \frac{c^2 \alpha^2}{\omega^2} \right)
%\end{equation}
%
Since $ E_{ z \omega } $ can be expressed through the $H_{ \varphi \omega } $
\begin{equation}
\label{Hphi2Ez}
E_{ z \omega } = \frac{ i }{ k_0 r} \pdv{ ( r H_{\varphi\omega} ) }{r},
\end{equation}
%
%Note that continuity of functions $ E_{ z \omega } $ and $ H_{ \varphi \omega } $ at $ r = a $, $ z > 0 $ leads to the equality $ \Phi_+^{ ( 1 ) } ( a, \alpha ) = \Phi_+^{ ( 2 ) } ( a, \alpha ) $.
%%
%Direct calculations with using \eqref{psi_1_sol} result in the following:
then we obtain:
\begin{equation}\begin{aligned}
\label{Phi_1_sol}
\Phi_+^{ ( 1 ) } ( r, \alpha )
= C_1 \frac{ i \kappa }{ k_0 } J_0 ( \kappa r ) + \\
\frac{ i }{ k_0 }\frac{ i }{ 2 \pi }
\left[
\frac{ \chi_l J_0 ( \chi_l r ) }{ k_{ z l } + \alpha } -
\sum\limits_{ m = 1 }^{ \infty } 
R_m \frac{ \chi_m J_0 ( \chi_m r ) }{ k_{ z m } - \alpha } 
\right],
\end{aligned}\end{equation}
where $k_0 = \omega / c$. 
Using the function $\Psi^{(2)} (r,\alpha)$ \eqref{psi_2_sol} it is also obtained
\begin{equation}
\label{Phi_2_sol}
\Phi^{ ( 2 ) } ( r, \alpha )
= 
C_2 \frac{ i \kappa }{ k_0 } H_0^{ ( 1 ) } ( \kappa r ).
\end{equation}
Coefficient $C_2$ can be determined using the continuity conditions at the external waveguide surface $r=a+0$, $z<0$. 
Recall that we consider the case of the smooth perfectly conductive external surface of the waveguide. 
Then $E_{z\omega} = 0$ at $r=a+0$, $z<0$ and therefore
$$
\Phi^{(2)}_-(a, \alpha ) = 0,
$$
and
\begin{equation}
\label{C2}
C_2 = \frac{ k_0 }{ i }\frac{ \Phi_+^{(2)} (a,\alpha) }{ \kappa H_0^{(1)} (\kappa a) }.
\end{equation}

\subsection{Wiener-Hopf-Fock equation}

Using \eqref{psi_2_sol} and \eqref{C2} one can write:
\begin{equation}
\label{WHFore}
\Psi_+^{ ( 2 ) } ( a, \alpha ) + \Psi_-^{ ( 2 ) } ( a, \alpha )
= \frac{ k_0 \Phi_+^{ ( 2 ) } ( a, \alpha )}{ i \kappa H_0^{(1)}(\kappa a )}
H_1^{(1)}(\kappa a ).
\end{equation}
To obtain Wiener-Hopf-Fock equation for the problem, one should express $\Psi_+^{ ( 2 ) } ( a, \alpha )$ through $\Phi_+^{ ( 2 ) } ( a, \alpha )$. This can be done using continuity conditions for $r=a$, $z>0$: 
\begin{equation}\begin{aligned}
\label{Phi_1_sol_a}
\Phi_+^{ ( 1 ) } ( a, \alpha )
= 
\Phi_+^{ ( 2 ) } ( a, \alpha ), \\
\Psi_+^{ ( 1 ) } ( a, \alpha )
= 
\Psi_+^{ ( 2 ) } ( a, \alpha ).
\end{aligned}\end{equation}
Excluding $C_1$ we obtain:
\begin{equation}
\begin{aligned}
\label{psi2phi2}
&\Psi_+^{ ( 2 ) } ( a, \alpha ) =
\frac{1}{J_0(\kappa a)}
\left[
\frac{k_0 J_1(\kappa a)}{ i \kappa } \Phi_+^{(2)}(a, \alpha) -
\right. \\
&-\frac{i}{2\pi}
\left(
\frac{ \frac{ J_1 (\kappa a) }{ \kappa } \chi_l J_0( a \chi_l ) - J_0(a\kappa)J_1(a\chi_l) }{ k_{zl}+\alpha} - 
\right.
\\ 
&-
\left.
\left.
\sum\limits_{m=1}^{\infty} R_m
\frac{ \frac{ J_1 (\kappa a) }{ \kappa } \chi_m J_0( a \chi_m ) - J_0(a\kappa)J_1(a\chi_m) }{ k_{zm}-\alpha} 
\right)  
\right].
\end{aligned}
\end{equation}
One can see that points
\begin{equation}
\label{alpham}
\alpha = \alpha_m \equiv \sqrt{k_0^2 - \left(\frac{j_{0m}}{a}\right)^2},
\quad
\mathrm{Im}\sqrt{}>0,
\end{equation}
are only singularities of the right-hand side of \eqref{psi2phi2},
$j_{0m}$ are zeros of the Bessel function $J_0$.
To eliminate these singularities (the left-hand side of \eqref{psi2phi2} is regular in the upper half-plane) one should require the following:
\begin{equation}
\label{exclude_pole}
\begin{aligned}
&\frac{a k_0 J_1(j_{0p}}{i j_{0p}}\Phi_+^{ ( 2 ) } ( a, \alpha_p ) = \\
&=\frac{ i }{ 2 \pi }
\left[
\frac{ \xi_l( \alpha_p ) }{ k_{ z l } + \alpha_p } 
- 
\sum\limits_{m=1}^{\infty}  
R_m \frac{ \xi_m ( \alpha_p ) }{ k_{ z m } - \alpha_p }
\right],
\end{aligned}\end{equation}
where $ p = 1,2, \ldots$,
\begin{equation}
\label{xim}
\xi_m ( \alpha ) = 
\frac{ J_1 (\kappa a) }{ \kappa } \chi_m J_0( a \chi_m ) - J_0(a\kappa)J_1(a\chi_m).
\end{equation}
Note that
\begin{equation}
\label{ximp}
\xi_m ( \alpha_p ) = 
\frac{a\chi_m J_1 ( j_{0p} ) J_0( a \chi_m ) }{ j_{0p} }.
\end{equation}
Substituting \eqref{psi2phi2} to \eqref{WHFore} one can obtain
\begin{equation}
\label{WHF-1}
\begin{aligned}
&0 = \frac{ 2 k_0 \Phi_+^{ ( 2 ) } ( a, \alpha ) }
{ \kappa( \alpha ) G( \alpha ) }
+\Psi_-^{ ( 2 ) }( \alpha ) - \\
&-
\frac{i}{2\pi}
\left[
\frac{ \xi_l( \alpha ) }{ J_0(\kappa a) ( k_{ z l } + \alpha ) } 
- 
\sum\limits_{m=1}^{\infty}  
\frac{R_m \xi_m ( \alpha ) }{ J_0( \kappa a) ( k_{ z m } - \alpha ) }
\right], 
\end{aligned}
\end{equation}
where
\begin{equation}
\label{G}
G(\alpha) = \pi a \kappa(\alpha)J_0(\kappa a)H_0^{(1)}(\kappa a ).
\end{equation}
This is Wiener-Hopf-Fock equation of the problem.
Performing standard factorization $\kappa(\alpha) = \kappa_+(\alpha)\kappa_-(\alpha)$, $\kappa_{\pm}(\alpha)=\sqrt{k_0\pm\alpha}$,
$G(\alpha) = G_+(\alpha)G_-(\alpha)$, and decomposing the following functions
\begin{equation}
\label{etazeta}
\begin{aligned}
\eta_{l}(\alpha) &=
\frac{\xi_l(\alpha) \kappa_-(\alpha) G_-(\alpha)}{J_0(\kappa a) ( k_{zl}+\alpha )}, \\
\zeta_{m}(\alpha) &=
\frac{\xi_m(\alpha) \kappa_-(\alpha) G_-(\alpha)}{J_0(\kappa a) ( k_{zm}-\alpha )},
\end{aligned}
\end{equation}
into sums $\eta_l(\alpha)=\eta_{l+}(\alpha)+\eta_{l-}(\alpha)$,
$\zeta_m(\alpha)=\zeta_{m+}(\alpha)+\zeta_{m-}(\alpha)$, where
\begin{equation}
\label{etazeta+}
\begin{aligned}
\eta_{l+}(\alpha) &=
-\frac{\chi_l J_0(a\chi_l)}{a}
\sum\limits_{n=1}^{\infty}
\frac{\kappa_+(\alpha_n)G_+(\alpha_n)}{\alpha_n(k_{zl}-\alpha_n)(\alpha_n + \alpha)}, \\
\zeta_{m+}(\alpha) &=
-\frac{\chi_m J_0(a\chi_m)}{a}
\sum\limits_{n=1}^{\infty}
\frac{\kappa_+(\alpha_n)G_+(\alpha_n)}{\alpha_n(k_{zm}+\alpha_n)(\alpha_n + \alpha)},
\end{aligned}
\end{equation}
we can rewrite the Wiener-Hopf-Fock equation \eqref{WHF-1} so that the left part contains only functions regular in area $\text{Im} \alpha > 0$ and the right part --- functions regular in area $\text{Im} \alpha < 0$:
\begin{equation}
\begin{aligned}
\label{WHF-2}
&\frac{ 2 k_0 \Phi_+^{ ( 2 ) } ( a, \alpha ) }
{ \kappa_+ ( \alpha ) G_+ ( \alpha ) } - \\
&-
\frac{i}{2\pi}
\left(
\eta_{l+}(\alpha) - \sum\limits_{ m = 1 }^{ \infty } R_m \zeta_{m+}( \alpha ) 
\right) = \\
&= - \Psi_-^{ ( 2 ) }( a, \alpha ) \kappa_- ( \alpha ) G_- ( \alpha ) + \\
&+ \frac{i}{2\pi}
\left(
\eta_{l-}(\alpha) - \sum\limits_{ m = 1 }^{ \infty } R_m \zeta_{m-}( \alpha )
\right) = P( \alpha ),
\end{aligned}
\end{equation}
where $ P( \alpha ) $ is a some polynomial.

Using the general theorems~%
\cite{Fock44} we can conclude that $ G_{ \pm } (\alpha) \to 1 $. 
Performing other typical estimations of functions asymtotic behavior in \eqref{WHF-2}, one can conclude that $P (\alpha) \sim \alpha^{-1} $ for $|\alpha| \to \infty $, therefore $ P(\alpha) = 0$. 
Thus, the formal solution of the Wiener-Hopf-Fock equation is constructed: 
\begin{equation}
\begin{aligned}
\label{WHF_sol}
&\Phi_+^{ ( 2 ) } ( a, \alpha ) =
\frac{ i \kappa_+( \alpha ) G_+( \alpha ) }{ 4 \pi k_0 } \times \\ 
&\times
\left(
\eta_{l+}( \alpha ) - \sum\limits_{ m = 1 }^{ \infty } R_m \zeta_{m+} ( \alpha )
\right).
\end{aligned}\end{equation}

\subsection{Reflection coefficients $ R_m $}

By substituting the solution \eqref{WHF_sol} into \eqref{exclude_pole} and combining terms with $R_m$ we obtain the following infinite system of equations 
\begin{equation}
\label{sys}
\sum\limits_{ m = 1}^{ \infty }
W_{ p m } R_m = w_p,
\end{equation}
where
\begin{equation}
\label{W}
\begin{aligned}
&W_{ p m }
=
\kappa_+( \alpha_p ) G_+( \alpha_p ) \zeta_{m+}( \alpha_p ) -
2i
\frac{\chi_m J_0(a \chi_m )}{k_{zm} - \alpha_p}, \\
&w_p
=
\kappa_+( \alpha_p ) G_+( \alpha_p ) \eta_{l+}( \alpha_p ) -
2i
\frac{\chi_l J_0(a \chi_l )}{k_{zl} + \alpha_p}.
\end{aligned}
\end{equation}

An important note should be done here. 
One possible solution of the dispersion equation~\eqref{disp_eq} is $\chi_m \to 0$.
Corresponding mode has its $k_{zm} \to k_0$ therefore its phase velocity is close to the light speed $c$.
Therefore such a mode will be excited, for example, by relativistic charged particle bunch.
   
To take into account this situation it is expedient to introduce new reflection coefficients $\tilde{R}_m$ so that $R_m = \tilde{R}_m / \chi_m$.
In this case we obtain, for example, for single reflected waveguide mode:
\begin{equation}
\label{reflectedtilde}
H_{\varphi \omega m}^{\left(r\right)} = \tilde{R}_m \frac{ J_1 ( \chi_m r )}{\chi_m} \exp\left( -i k_{zm} z \right),
\end{equation}
which is finite for $\chi_m \to 0$.
Similar substitutions can be easily done in all appropriate cases, finally we obtain a modified system for $\tilde{R}_m$:
\begin{equation}
\label{systilde}
\sum\limits_{ m = 1}^{ \infty }
\tilde{W}_{ p m } \tilde{R}_m = \tilde{w}_p,
\end{equation}
where
\begin{equation}
\label{Wtilde}
\begin{aligned}
&\tilde{W}_{ p m }
=
\kappa_+( \alpha_p ) G_+( \alpha_p ) \frac{ \zeta_{m+}( \alpha_p ) }{ \chi_m } -
2i
\frac{ J_0(a \chi_m )}{k_{zm} - \alpha_p }, \\
&\tilde{w}_p
=
\kappa_+( \alpha_p ) G_+( \alpha_p ) \frac{ \eta_{l+}( \alpha_p ) }{ \chi_l }-
2i
\frac{ J_0(a \chi_l )}{k_{zl} + \alpha_p}.
\end{aligned}
\end{equation}
As a result, the unknown coefficients $R_m$ or $\tilde{R}_m$ can be easily calculated using the reduction method. 

%%%%%%%%%%%%%%%%%%%%%%%%%%%%%%%%%%%%%%%%%%%%%%%%%%%
\section{The case of uniform dielectric filling}
%%%%%%%%%%%%%%%%%%%%%%%%%%%%%%%%%%%%%%%%%%%%%%%%%%%

\begin{figure}
\centering
\includegraphics[width = 0.9\linewidth]{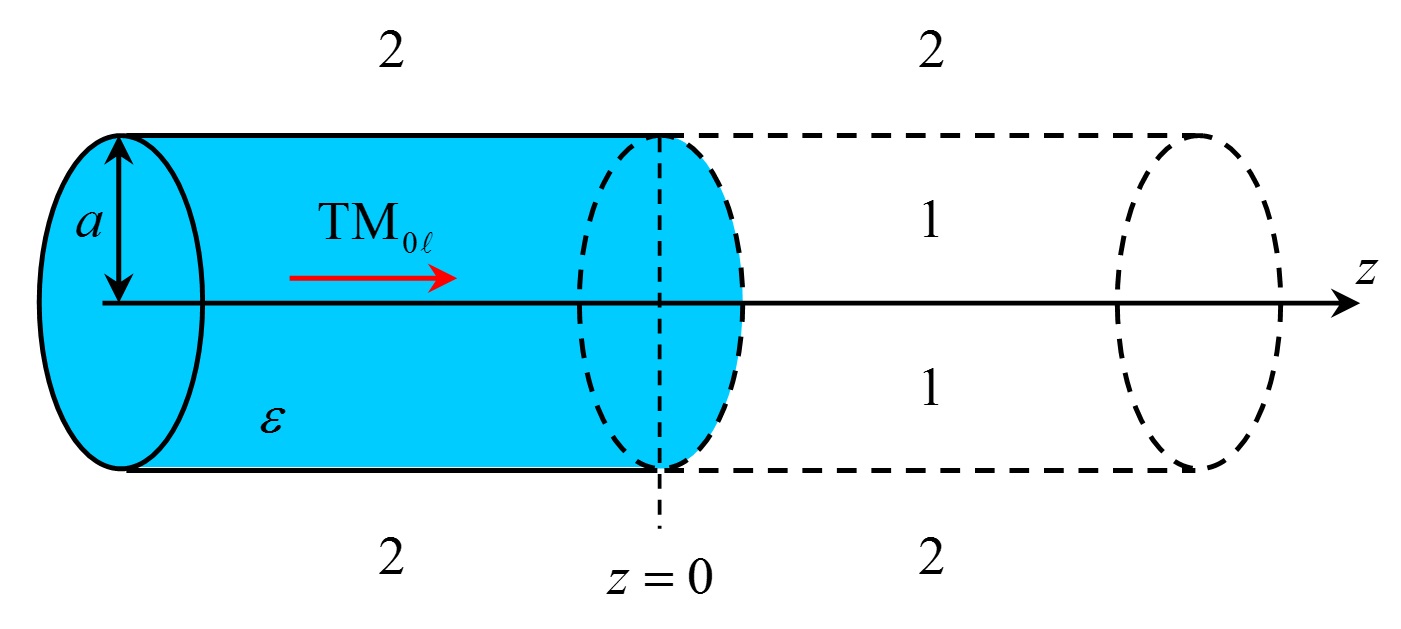}
\caption{\label{waveguided}Geometry of the problem for the case of uniform dielectric filling. Boundary conditions~\eqref{cor_con} are applied inside the waveguide while the outside waveguide wall is smooth and perfectly conductive. Dashed lines denote the imaginary waveguide continuation dividing the outer space into subregions ``1'' and ``2''. }
\end{figure}

For the case where waveguide is filled with homogeneous dielectric with permittivity $\varepsilon$ (see Fig.~\ref{waveguided}), the equivalent boundary conditions (EBC) have similar form~\eqref{cor_con} while impedance is slightly modified~\cite{TVAA2018}:
\begin{equation}
\label{impedanced}
\eta_m^d ( \varkappa_m ) = i k_0 \sqrt{\varepsilon} \left( \frac{d_2 d_3}{d} - \delta \frac{ \varkappa_m^2 }{ k_0^2 } \right),
\end{equation}
where longitudinal wavenumbers $\varkappa_m$ are equal to 
\begin{equation}
\label{kzd}
\varkappa_m = \sqrt{ k_0^2\varepsilon - \left(\chi_m^d \right)^2 },
\quad
\mathrm{ Im }  (\varkappa_m) > 0,
\end{equation}
while the transverse wavenumbers $ \chi_m^d $ are determined by the dispersion equation
\begin{equation}
\label{disp_eqd}
J_0 \left( \chi_m^d a \right)- k_0\varepsilon | \eta_m^d | \frac{ J_1 \left( \chi_m^d a \right) }{ \chi_m^d } = 0.
\end{equation}
Again, we can suppose that $\{\chi_m^d\}$ are known since they can be found numerically for arbitrary set of specific problem parameters.

An incident mode (with number $l$) and reflected field decomposed into a series of reflected eigenmodes are expressed as follows:
\begin{equation}
\label{initiald}
H_{\varphi \omega}^{\left(i\right)d} = \frac{ J_1 \left( \chi_{l} r \right) }{ \chi_{l} } \exp\left( i \varkappa_l z \right),
\end{equation}
\begin{equation}
\label{reflectedd}
H_{\varphi \omega}^{\left(r\right)d} = \sum\limits_{m=1}^{\infty} R_m^d 
\frac{ J_1 \left( \chi_{m} r \right) }{ \chi_{m} } \exp\left( -i \varkappa_m z \right),
\end{equation}
while the rest of components is calculated as follows:
\begin{equation}
\label{othercompd}
\begin{aligned}
&E_{r\omega} = \frac{ 1 }{ i k_0 \varepsilon } \pdv{ H_{\varphi \omega} }{z}, \\
&E_{z\omega} = \frac{ i }{ k_0 \varepsilon } \frac{1}{r} \pdv{\left( r H_{\varphi \omega} \right) }{r}.
\end{aligned}
\end{equation}

Further steps are very similar to those for vacuum case discussed above.
Moreover, they are also close to those from the problem with an open-ended dielectric-filled waveguide with smooth walls both inside and outside~\cite{GVT2021}.
Therefore below we shortly mention only the main steps of the solution (similar functions and constants related to the case of dielectric filling will be marked by the superscript ``d'').

So, the solutions for integral transforms have the form 
\begin{equation}
\begin{aligned}
\label{psi_1_sold}
\Psi_+^{(1)d} (r,\alpha) &= C_1^d J_1 ( \kappa r ) + \frac{i}{2\pi} \frac{ J_1 ( \chi_l^d r ) }{ \chi_l^d }
\frac{\frac{ \varkappa_l }{\varepsilon} - \alpha }{  k_{ z l }^2 - \alpha^2 } - \\
&- \frac{i}{2\pi} 
\sum\limits_{m=1}^{\infty} R_m^d \frac{ J_1 ( \chi_m^d r ) }{ \chi_m^d }
\frac{ \frac{ \varkappa_m }{ \varepsilon } + \alpha }{ k_{zm}^2 - \alpha^2 },
\end{aligned}
\end{equation}
\begin{equation}
\Psi^{ ( 2 )d } ( r, \alpha ) = C_2^d H_1^{(1)} (\kappa r),
\label{psi_2_sold}
\end{equation}
\begin{equation}
\begin{aligned}
\label{Phi_1_sold}
&\Phi_+^{ ( 1 )d } ( r, \alpha )
= C_1^d \frac{ i \kappa }{ k_0 } J_0 ( \kappa r ) + \frac{ i }{ k_0 }\frac{ i }{ 2 \pi } \times\\
&\times
\left[
J_0 ( \chi_l^d r )\frac{ \frac{ \varkappa_l }{ \varepsilon } - \alpha }{ k_{ z l }^2 - \alpha^2 } -
\sum\limits_{ m = 1 }^{ \infty } 
R_m^d J_0 ( \chi_m r ) \frac{ \frac{ \varkappa_m }{ \varepsilon } + \alpha  }{ k_{ z m }^2 - \alpha^2 } 
\right],
\end{aligned}
\end{equation}
\begin{equation}
\label{Phi_2_sold}
\Phi^{ ( 2 )d } ( r, \alpha )
= 
C_2^d \frac{ i \kappa }{ k_0 } H_0^{ ( 1 ) } ( \kappa r ).
\end{equation}

Coefficient $C_2^d$ is found as usual,
\begin{equation}
\label{C2d}
C_2^d = \frac{ k_0 }{ i }\frac{ \Phi_+^{(2)d} (a,\alpha) }{ \kappa H_0^{(1)} (\kappa a) }.
\end{equation}
and after excluding $C_1^d$ from \eqref{psi_1_sold} and \eqref{Phi_1_sold} we obtain:
\begin{equation}
\begin{aligned}
\label{psi2phi2d}
&\Psi_+^{ ( 2 )d } ( a, \alpha ) =
\frac{ 1 }{ J_0( \kappa a ) }
\left[
\frac{ k_0 J_1(\kappa a) }{ i \kappa } \Phi_+^{ ( 2 )d }( a, \alpha ) -
\right. \\
&-\frac{i}{2\pi}
\frac{ \frac{ \varkappa_l }{ \varepsilon } - \alpha }{ k_{zl}^2 - \alpha^2 }
\left(
\frac{ J_1 (\kappa a) }{ \kappa } J_0( a \chi_l^d ) - J_0(a\kappa) \frac{ J_1( a\chi_l^d) }{ \chi_l^d }
\right) - \\ 
&-
\frac{i}{2\pi}\sum\limits_{m=1}^{\infty} R_m
\frac{ \frac{ \varkappa_m }{ \varepsilon } + \alpha }{ k_{zm}^2 - \alpha^2 } \times \\
&\times
\left.
\left(
\frac{ J_1 (\kappa a) }{ \kappa } J_0( a \chi_m ) - J_0(a\kappa) \frac{ J_1( a \chi_m^d ) }{ \chi_m^d }
\right) 
\right].
\end{aligned}
\end{equation}
``Regularization condition'', similar to \eqref{exclude_pole}, has the form
\begin{equation}
\label{exclude_polesd}
\begin{aligned}
&\frac{a k_0 J_1(j_{0p}}{i j_{0p}}\Phi_+^{ ( 2 )d } ( a, \alpha_p ) = \\
&=\frac{ i }{ 2 \pi }
\left[
\phi_l( \alpha_p )
\frac{ \frac{ \varkappa_l }{ \varepsilon } - \alpha_p }{ k_{ z l }^2 - \alpha_p^2 } 
- 
\sum\limits_{m=1}^{\infty}  
R_m^d \phi_m ( \alpha_p )
\frac{ \frac{ \varkappa_m }{ \varepsilon } + \alpha_p }{ k_{ z m }^2 - \alpha_p^2 }
\right],
\end{aligned}
\end{equation}
where $ p = 1,2, \ldots$,
\begin{equation}
\label{phim}
\phi_m ( \alpha ) = 
\frac{ J_1 (\kappa a) }{ \kappa } J_0( a \chi_m^d ) - J_0(a\kappa) \frac{ J_1( a \chi_m^d ) }{ \chi_m^d }.
\end{equation}

Wiener-Hopf-Fock equation of the problem (after standard decompositions $\kappa(\alpha) = \kappa_+(\alpha)\kappa_-(\alpha)$, 
$G(\alpha) = G_+(\alpha)G_-(\alpha)$) takes the form:
\begin{equation}
\label{WHF-1d}
\begin{aligned}
&0 = \frac{ 2 k_0 \Phi_+^{ ( 2 )d } ( a, \alpha ) }
{ \kappa_+( \alpha ) G_+( \alpha ) }
+\Psi_-^{ ( 2 )d }( \alpha )\kappa_-( \alpha ) G_-( \alpha ) - \\
&-
\frac{i}{2\pi}
\left[
\varphi_l( \alpha ) -
\sum\limits_{m=1}^{\infty}  
R_m^d \psi_m ( \alpha ) \right], 
\end{aligned}
\end{equation}
where
\begin{equation}
\label{phipsid}
\begin{aligned}
\varphi_l( \alpha ) 
&= 
\frac{ \frac{ \varkappa_l }{ \varepsilon } - \alpha }{ k_{ z l }^2 - \alpha^2 } 
\frac{ \phi_l( \alpha ) }{ J_0( \kappa a ) }
\kappa_-(\alpha) G_-(\alpha), \\
\psi_m( \alpha ) 
&= 
\frac{ \frac{ \varkappa_m }{ \varepsilon } + \alpha }{ k_{ z m }^2 - \alpha^2 } 
\frac{ \phi_m( \alpha ) }{ J_0( \kappa a ) }
\kappa_-( \alpha ) G_-( \alpha ).
\end{aligned}
\end{equation}
Formal solution of the Wiener-Hopf-Fock equation is the following: 
\begin{equation}
\begin{aligned}
\label{WHF_sold}
&\Phi_+^{ ( 2 )d } ( a, \alpha ) =
\frac{ i \kappa_+( \alpha ) G_+( \alpha ) }{ 4 \pi k_0 } \times \\ 
&\times
\left(
\varphi_{l+}( \alpha ) - \sum\limits_{ m = 1 }^{ \infty } R_m^d \psi_{m+} ( \alpha )
\right),
\end{aligned}
\end{equation}
where $\varphi_{l+}( \alpha )$ and $\psi_{m+}( \alpha ) $ are ``+'' summands of $\varphi_l( \alpha )$ and $\psi_m( \alpha )$, respectively.
Being substituted to~\eqref{exclude_polesd}, solution~\eqref{WHF_sold} results in the following infinite system of linear equations for 
$R_m^d$  
\begin{equation}
\label{sysd}
\sum\limits_{ m = 1}^{ \infty }
W_{ p m }^d R_m^d = w_p^d,
\end{equation}
where
\begin{equation}
\label{Wd}
\begin{aligned}
&W_{ p m }^d
=
J_0( a \chi_m^d )
\left[
2ia\frac{ \frac{ \varkappa_m }{ \varepsilon } + \alpha_p }{ k_{ z m }^2 - \alpha_p^2 } + \right. \\
&+
\left.
\kappa_+( \alpha_p ) G_+( \alpha_p ) 
\sum\limits_{q=1}^{\infty}
\frac{ \kappa_+( \alpha_q ) G_+( \alpha_q ) }{ k_{zm}^2 - \alpha_q^2 }
\frac{ \frac{ \varkappa_m }{ \varepsilon } - \alpha_q }{ ( \alpha_p + \alpha_q ) \alpha_q }
\right], \\
&w_p^d
=
J_0( a \chi_l^d )
\left[
2 i a \frac{ \frac{ \varkappa_l }{ \varepsilon } - \alpha_p }{ k_{ z l }^2 - \alpha_p^2 } + \right. \\
&+
\left.
\kappa_+( \alpha_p ) G_+( \alpha_p ) 
\sum\limits_{q=1}^{\infty}
\frac{ \kappa_+( \alpha_q ) G_+( \alpha_q ) }{ k_{zl}^2 - \alpha_q^2 }
\frac{ \frac{ \varkappa_l }{ \varepsilon } + \alpha_q }{ ( \alpha_p + \alpha_q ) \alpha_q }
\right].
\end{aligned}
\end{equation}
Therefore, the problem is solved.

It should be noted that system~\eqref{sysd} is similar in form to corresponding system for the problem with dielectric-lined open-ended waveguide with smooth perfectly conducting walls~\cite{GV2021arxiv}. 
In~\cite{GV2021arxiv}, corresponding analytic solution was used to perform a series of numerical calculations based on reduction of infinite system to a finite one.
Moreover, a comparison with simulations in COMSOL has been performed and an excellent agreement has been observed.
Thus no problems are expected with numerical solution of~\eqref{sys} and \eqref{sysd} using the reduction technique.
On the contrary, numerical simulation of these problems in, say, COMSOL is seemed to be impossible because it is impossible to directly incorporate the equivalent boundary conditions (EBC)~\eqref{cor_con} into simulations with semi-infinite waveguide.
It should be underlined that~\eqref{cor_con} are principally different from the ordinary impedance boundary conditions available in COMSOL -- they are anisotropic and operate with purely imaginary impedance.    
Thus, the realistic scenario for simulations is to consider a section of smooth waveguide before the section with shallow corrugation.
Standard mode of a smooth section can be easily excited using COMSOL instruments, and this mode in turn will excite the corrugated section with open end.
However, analytic solution of such a problem is much more complicated.    

\begin{acknowledgments}
This work was supported by Russian Science Foundation (Grant No. 18-72-10137).
\end{acknowledgments}
    
%\bibliography{SNG_Bibliography_Apr2021}
%

\end{document}